\documentclass[%
 reprint,
groupedaddress,
 amsmath,amssymb,
prb,
]{revtex4-1}

\usepackage[final]{graphicx}
\usepackage{dcolumn}
\usepackage{bm}

\begin{document}


\title{Contrasting Elastic Properties of Heavily B- and N-doped Graphene,
with Random Distributions Including Aggregates}

\author{Karolina Z. Milowska}
 \affiliation{ Institute of Theoretical Physics, Faculty of Physics, University of Warsaw, ul. Ho\.za 69,
PL-00-681 Warszawa, Poland}
\author{Magdalena Woi\'nska}%
\affiliation{Faculty of Chemistry, University of Warsaw, ul. Pasteura 1,
PL-02-093 Warszawa, Poland}
\author{Ma\l gorzata Wierzbowska}
\affiliation{Institute of Theoretical Physics, Faculty of Physics, University of Warsaw, ul. Ho\.za 69,
PL-00-681 Warszawa, Poland}
\email{malgorzata.wierzbowska@fuw.edu.pl}
\date{\today}

\begin{abstract}
We focused on elastic properties of B- and N-doped graphene in wide range of concentrations
up to 20\%. The Young's, bulk and shear moduli and Poisson's ratio have been calculated
by means of the density functional theory for a representative set of supercells with
disordered impurity patterns including aggregates.  
In contrast to earlier work, it is demonstrated that doping with nitrogen even strengthens
the graphene layers, whereas incorporation of boron induces large structural and 
morphological changes seen in simulated STM images. Young's and shear moduli 
increase or decrease with the doping strength for nitrogen or boron, respectively,
while bulk modulus and Poisson's ratio exhibit opposite trends. Elastic properties
of samples for both types of impurities are strongly related to the electronic structures, 
especially for heavy doping ($>$12\%). Local arrangements of dopants and an agregation
or separation of impurities play crucial role in the determination of stiffness in 
the investigated systems. Interestingly, these findings are opossed for 
B- and N-contained samples. 
\end{abstract}
\maketitle

\section{Introduction}
Nowadays graphene, a two-dimensional monolayer formed out of sp$^2$ hybridizated carbon 
atoms ordered in a honeycomb-like lattice, attracts a lot of attention owing to its 
unique properties. Due to the hexagonal symmetry, 
its valence and conduction bands cross and have linear dispersion at high symmetric K-point;
 these bands determine semimetallic character of graphene and its extremely high electron 
mobility.\cite{geim, georgakilas} This, in connection with excellent mechanical properties, 
renders graphene to be an ideal candidate for applications in flexible 
electronics\cite{bunch,standley,pereira,guinea,levy,lu2012} and nanocomposites.\cite{stankovich, ramanathan} 
However, the ability of generating controllable band gap in graphene is a prerequisite 
for effective applications in transistor based electronic devices.\cite{schwiertz} 
Therefore, an effective functionalization that would open the zero energy band gap 
in pristine graphene without significant deterioration of the remaining advantageous 
properties is searched for, and practically, any thinkable way  of reaching this target is 
currently investigated. Substitutional doping, which refers to substitution of carbon atoms 
in graphene lattice by atoms with different number of valence electrons such as boron or 
nitrogen, could be a recipe to achieve this goal.\cite{georgakilas,liu} 

Compared to carbon, nitrogen has one additional electron and boron lacks one, 
which means that these elements should act as electron donors and acceptors, respectively. 
Despite the poor environmental stability of chemically doped graphene,\cite{wu}  
there are many experimental and theoretical studies that show a possibility to 
prepare p- and n-type semiconductors by substituting these elements into 
graphene.\cite{liu,mukherjee,arxiv,appa,yuge,hanafusa} 
It has been previously shown that, the Fermi level shifts with respect to 
the valence band top or the conduction band bottom resulting from B or N 
functionalization, correspondingly, and the band gap 
opens.\cite{arxiv,wei,mukherjee,wu,panchakarla2009,kaloni}. 
Boron doped graphene (BG) has been successfully synthesized using arc-discharge 
of graphite electrodes in the presence of H$_2$, He and 
B$_2$H$_6$\cite{panchakarla2009, panchakarla2010} and by the chemical vapor deposition (CVD) 
using  polystyrene and HBO$_3$.\cite{wu}
Whereas, single- and a few layer nitrogen doped graphene (NG) have been synthesized by the CVD 
with NH$_3$\cite{wei,zhao} or  CH$_4$N$_2$O\cite{wu} as nitrogen source and using the 
arc-discharge method in the presence of H$_2$ and C$_5$H$_5$N\cite{panchakarla2009} or H$_2$ 
and NH$_3$.\cite{panchakarla2009,panchakarla2010} 

The BG and NG materials open various potential applications like in graphene-based back-gate 
field-effect transistors,\cite{wu} nanosensors,\cite{martins,wang} fuel cells,\cite{qu} 
lithium ion batteries\cite{wang2,gao} or even hydrogen storage.\cite{beheshti} 
Therefore understanding the influence of substitutional doping, with boron and nitrogen, 
on graphene properties is proved to be crucial. 

In contrast to other authors, we find that the boron-doped graphene shows opposite elastic
properties with respect to the nitrogen-doped graphene. These effects are especially 
well pronounced in heavily doped samples, not studied theoretically so far
but extensively used in tailoring electronic devices\cite{georgakilas}. 
The effects of dopants aggregation and disorder play a huge role
in the determination of Young's, bulk and shear moduli as well as the Poisson's ratio.    
In this work, we analyze series of arrangements of impurities in the graphene layer,
starting from small concentrations and ending with 20\%.   
Therefore, we extended our previous theoretical studies,\cite{arxiv} on electronic properties
of these systems, to mechanical and structural behaviour at the parameters close to 
the technological conditions. 

\section{Theoretical methods}

Our studies of functionalized graphene layer are based on {\it ab initio} 
calculations within the framework of the spin polarized density functional 
theory (DFT).\cite{hohenberg,kohn} Generalized gradient approximation of 
the exchange-correlation functional in Perdew-Burke-Ernzerhof (PBE) 
parameterization has been applied.\cite{perdew} Calculations have been 
performed using the SIESTA package.\cite{portal,soler} 
Valence electrons have been represented with double zeta basis sets of 
orbitals localized on atoms, with polarization functions also included. 
The influence of core electrons has been accounted within the pseudopotential 
formalism. Norm-conserving Troullier-Martins nonlocal 
pseudopotentials\cite{troullier} used in our studies were cast into 
the Kleinman-Bylander separable form.\cite{kleinman} The energy 
cut-off for the density on the real space grid has been set to 800 Ry. 
The Brillouin zone has been sampled in the 5x5x1 
Monkhorst and Pack scheme.\cite{mp}  
Calculations were performed within the supercell scheme with the graphene 
layers separated by a distance of 80 \AA, 
large enough to eliminate any interaction. 
Structural optimization has been conducted using the conjugate gradient 
algorithm to achieve the residual forces, acting on the atoms, 
lower than 0.001 eV/\AA.

We performed calculations for 5x5 supercells containing 25 graphene primitive 
unit cells (i.e., consisting of 50 atoms). Such supercell has been chosen 
in order to examine a wide range of dopants concentrations and a variety of 
possible distributions of the substituent atoms. In the described graphene 
supercell, one to ten B or N atoms have been introduced, leading to the 
corresponding concentration of 2-20$\%$. In the case of two substituted 
atoms in a supercell, all eleven symmetrically nonequivalent configurations 
of atoms have been examined. In the remaining cases, twelve different 
randomly chosen configurations have been taken into account.
For each sample at given concentration, and with unique dopants distribution
nonequivalent to other samples, the elastic properties have been calculated
and averaged at the end. The statistical spreads of obtained values, 
due to disorder and aggregation, are also presented.  

\begin{figure*}
\includegraphics[width=0.85\textwidth]{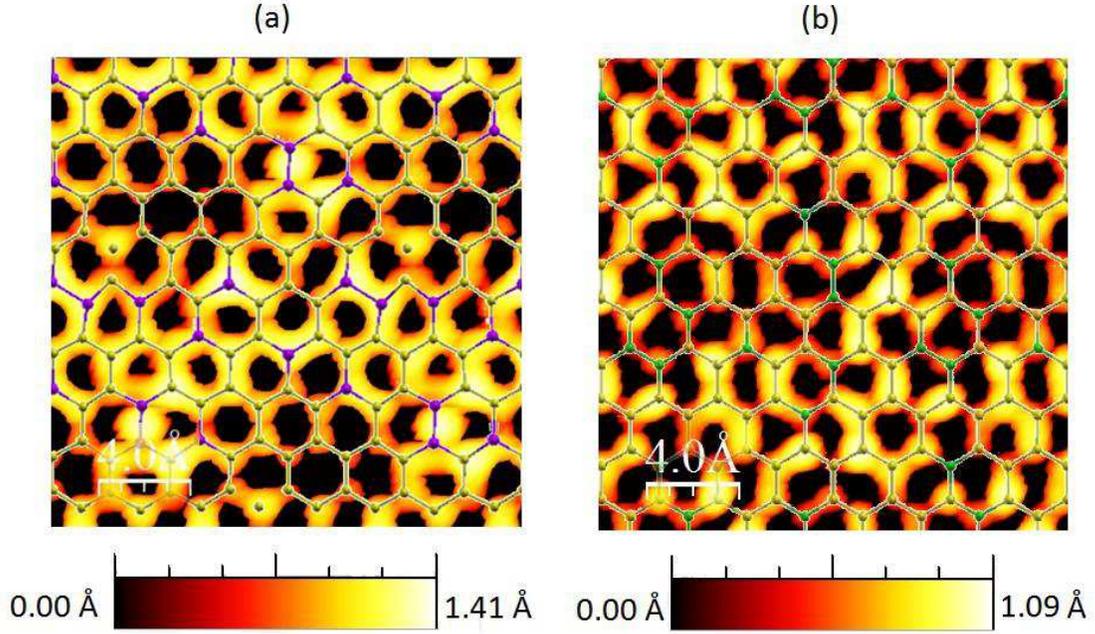}
\caption{\label{fig:stm}
Simulated STM images of graphene functionalized with B (a) and N (b) atoms
are visualised above (V$_{bias}$=0.5V, 20\% concentration of dopants). Superimposed is a ball-and-stick model
of the functionalized graphene lattice. The C atoms are depicted in yellow color,
B in purple and N in green.}
\end{figure*}

\begin{figure*}
\includegraphics[width=0.9\textwidth]{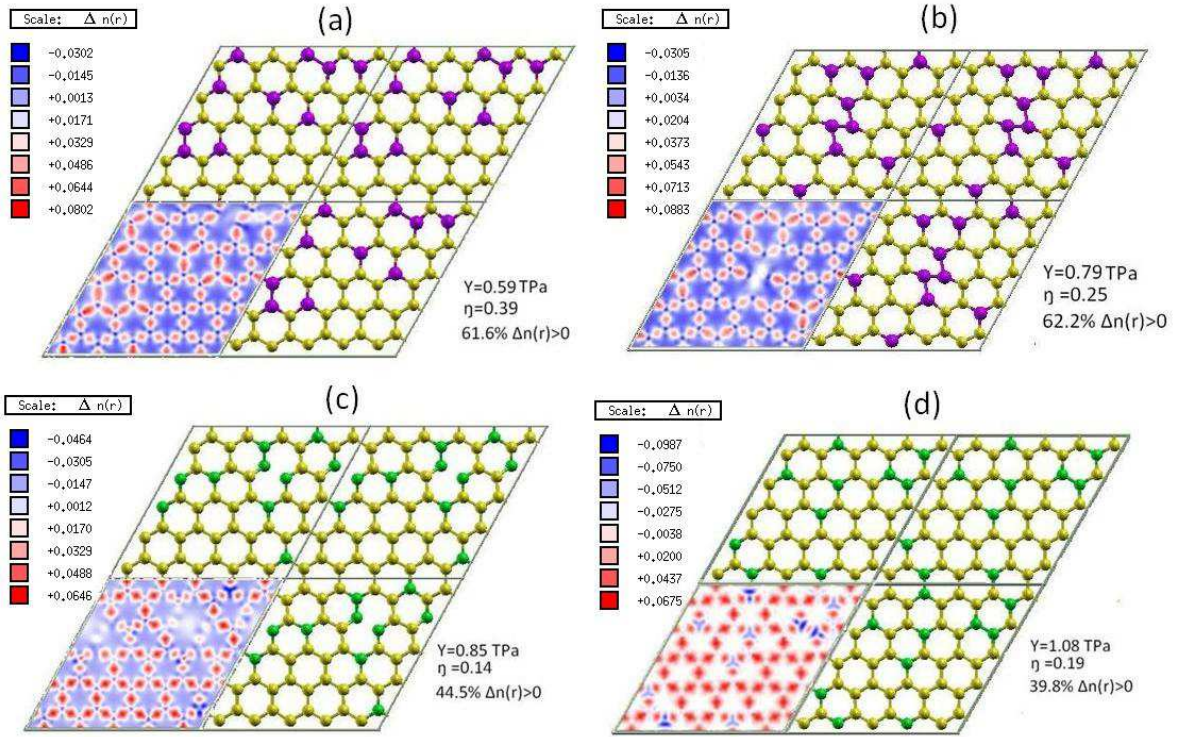}
\caption{\label{fig:drho} The  ball-and-stick model of four example structures of
graphene functionalized boron (a,b) and nitrogen atoms (c,d) presented in three (5x5) supercells.
In the bottom-left supercell the difference of the valence pseudocharge density and
the superposition of spherical atomic valence pseudocharge densities (DRHO) is depicted.
Concentration of substituent atoms: 20\%.}
\end{figure*}

We simulated the scanning tunneling microscopy (STM) images in order to 
analyze the doped graphene structures.
These images, depicted in Figure~\ref{fig:stm}, are calculated within 
the Tersoff-Hamann scheme\cite{Tersoff} in the constant current mode with the 
bias voltage (V$_{bias}$) of 0.5 V, 
and visualized using WSxM code.\cite{Horcas} 
Tip shape and its electronic structure were not taken into account. 
Density of the tunneling current, resulting from the applied small bias voltage 
between the tip and the sample, is calculated according to the 
formula:\cite{zheng}
\begin{eqnarray}
 \rho _{STM} (\vec r,V_{bias})  =  && \nonumber \\
\int\limits_{E_F  - eV_{bias}}^{E_F } \sum\limits_i^{N_{band}} 
\sum\limits_{\vec k \in BZ} 
| \varphi _{i,\vec k} (\vec r) |^2 \delta (E_{i,\vec k}  - E)dE, & &
\end{eqnarray}
where $\varphi _{i,\vec k}$ is the wave function associated with the 
$E_i$ eigenvalues from the energetic window set by the integrand range 
which contain $N_{band}$ states.

We obtained the elastic properties for all considered concentrations and 
symmetrically nonequivalent configurations of substituent atoms. 
All values for each concentration have been averaged over the investigated 
geometric configurations and compared with the relevant results for B or N 
substitutions. 
We constrained six different stress tensors with the nonzero component along the one of the lattice vectors to each system. The structures were optimized at the requested stress. At the end the obtained strain components were always smaller than 1\%.
The most interesting quantity, namely the Young's modulus, has been determined 
on the basis of the stress tensor $\sigma_{ii}$ and 
the strain $\varepsilon _{ii}$ components as follows:
\begin{equation}
\label{eyoung}
Y = \frac{{\sigma _{ii} }}{{\varepsilon _{ii} }}.
\end{equation}
The volume of doped systems in the supercell was calculated using the equation
\begin{equation}
\label{evolume}
V_o  = a \cdot b \cdot \sin (\theta ) \cdot 2 \; 
\frac{{r_S n_s  + r_C (n - n_s )}}{n}
\end{equation} 
where $a$ and $b$ are lattice constants of the 5x5 supercell, $\theta$ is the angle between them, 
$r_C$ and $r_S$ are van der Waals radii of C atom  (equal to 0.17 nm) 
and substituents (equal to 0.155 nm for N and 0.192 nm for B). The numbers 
$n_S$ and $n$ count dopant atoms and all the atoms in the supercell, respectively. 
The Poisson's ratio was obtained according to the formula	
\begin{equation}
\label{epoisson}
\eta  =  - \frac{{\Delta a}}{a}\;\frac{b'}{{\Delta b'}},
\end{equation}	
where  $b'$ is projection of the lattice vector $b$ on the direction perpendicular to lattice vector $a$, $\Delta a$ and $\Delta b'$ denote a change of  $a$ and $ b'$ in the strained graphene layer, respectively.
Bulk modulus (K) and shear modulus (G) (which could be easily derived from 
Eq.~\ref{eyoung} and Eq.~\ref{epoisson}) are given, respectively, by 
\begin{equation}
\label{ebulk}
K  =  \frac{Y}{3(1-2 \nu)},
\end{equation}
\begin{equation}
\label{eshear}
G  =  \frac{Y}{2(1+ \nu)},
\end{equation}
to complete a set of the elastic constants calculated in this work. 

\section{Results and discussion}

\subsection{Atomic and charge-density changes induced by doping}

\begin{table*}
\caption{Elastic properties of pure graphene and h-BN monolayers:
Young's (Y), bulk (K) and shear modulus (G) [in TPa], and Poisson's ratio ($\nu$),
in comparison to the experimental and other theoretical values. Some papers give
in-plane stiffness instead of Young's modulus, therefore, we have recalculated their
results using as the layer thickness the Van der Waals radius of C atom for pure graphene
or the averaged radius of N and B atoms for h-BN.}
\begin{tabular}{lcp{3cm}ccp{3cm}cc}
\hline
\hline \\[-0.2cm]
 Property & \multicolumn{3}{c}{graphene} & \multicolumn{3}{c}{h-BN} \\
   &    this work  &  calculated  &  exp. 
   &    this work  & calculated  &  exp.  \\[0.1cm]
\hline \\[0.1cm]
Y  & 1.050 &
     0.965\cite{topsakal} 0.994\cite{kudin} &
     0.980\cite{lee} 1\cite{koenig}  &
     0.756 &
     0.769\cite{topsakal} 0.781\cite{kudin} 0.802\cite{peng} 0.829\cite{mirzenhard}
     &    \\
K  & 0.528 &   &  0.582\cite{nag}  & 0.443 &   &  0.461\cite{nag} \\
G  & 0.449 &
     0.432\cite{kudin} 0.445\cite{zakharchenko} 0.366-0.460\cite{min} &
     0.280\cite{liu2012}  &
     0.311 &  0.323\cite{kudin} &  0.055\cite{wittkowski} \\
$\nu$   & 0.169 &
          0.149\cite{kudin} 0.16\cite{topsakal} &   &
          0.216 &
          0.21\cite{topsakal} 0.211\cite{kudin} 0.213\cite{mirzenhard} 0.217\cite{peng} &
            \\[0.1cm]
\hline
\hline
\end{tabular}
\label{tab:tab1}
\end{table*}

In our previous investigations,\cite{arxiv, appa} 
we focused on studies of the stability and electronic structure of B- and N-doped graphene. 
We found that these structures are stable, therefore it is interesting to proceed with 
deeper insight into physical and chemical properties.
Important conclusion from our earlier work was that substitution with boron atoms
is energetically more favorable than substitution with nitrogen atoms, which is consistent
with other theoretical work by Berseneva et al.\cite{berseneva}. We have also shown how
functionalization induces morphological changes in these systems and leads to
a redistribution of the electronic charge. 
Moreover, incorporation of foreign atoms not only destroys the sp$^2$ hybridization of 
the carbon atoms, but also causes deformations perpendicular to the graphene sheet.
Boron atoms, in comparison to nitrogen impurities, have a greater influence on 
the geometry of doped graphene layer. Mostly, it is due to the fact that the covalent
radius of boron is larger and that of nitrogen is similar to the radius of carbon. 
We have found that the graphene monolayer functionalized with boron atoms is no longer flat, 
and B atoms stick out from the surface. Similar observation has been also reported by other 
authors,\cite{panchakarla2009,gao,wu} and
this effect is better pronounced for higher concentrations of substituent atoms.
The specified changes noticed in the morphology of functionalized graphene,
their effect on the properties, and finally the wide range of potential applications,
encouraged presented here studies of the elastic moduli.

Before we turn to the discussion of the elastic properties, we present shortly
the structural and related electronic properties caused by the functionalization.
Scanning tunneling microscopy (STM) is a very powerful experimental 
technique which allows to investigate simultaneously the electronic and atomic structure
of samples. Therefore, we supplement our investigations with simulated STM images, which are 
presented in Figure~\ref{fig:stm} for graphene doped with boron (a) and nitrogen (b) 
at the bias voltage of 0.5 V. They provide useful information on the electron density of the 
occupied states.
For high doping concentration of 20\%, both images show deformed graphene structures,
where impurities change the bond lengths and angles in lattice. 
We see dark holes corresponding to small current in the hexagon centers, 
which are very characteristic for the pure graphene\cite{amara,zhao} and originate from 
destructive interference between tip and C atoms.
Boron-doped graphene presents large bright pattern centered on the B atoms, whereas nitrogen doped 
graphene is characterized by darker areas surrounding substituents. 
In the STM images, carbon atoms are brighter than nitrogen because the local charge density of 
N, at the distances pictured by this technique, has smaller contribution to $\rho_{STM}$ 
than that originating from the C atoms surrounding the impurities. 
At the same time in B-doped graphene, the opposite situation resulting from 
the occupied antibonding states of both boron and its first C-neighbours can be observed.
Similar pattern was obtained by Zheng\cite{zheng}. 
It is worth mentioning, that our simulated STM images of the N-doped graphene correlate well 
with Zhao's STM mapping.\cite{zhao} 

From a picture for the differential charge-density (DRHO) shown in Figure~\ref{fig:drho} (insets), 
it is clear how the electronic structure changes under particular arrangements of impurities. 
If the dopants are more clustered, as in Figure~\ref{fig:drho}(b) and (c), 
then the charge distribution is modified in larger areas than in the case of sparsely
distributed substituents, as in Figure~\ref{fig:drho}(a) and (d). 
Characteristic hexagons filled by negative differential charge-density and separated by 
positive differential density, which overlaps with bonds in most of the areas, cannot be clearly 
distinguished in the regions of clustered dopants. 
In the B-doped graphene, we observe the depletion of electron density in the B atoms vicinity 
and a transfer of the electronic charge to dopants and their first C-neighbors.

Due to one additional electron in N atoms as compared to carbon, in the N-doped graphene
the electronic transfer from dopants to the C-rings is observed.
In other words, N substituents induce strong intervalley electron scattering.
Charge transfer between both types of substituents and the C atoms was also observed in some
experiments\cite{panchakarla2009,zhao} and theoretical works.\cite{wang,zheng} 
The differential charge-density of the doped system is much more locally affected 
by the presence of B atoms than by N atoms, which is consistent with the STM analysis.

\subsection{Elastic properties of pure graphene and  hexagonal boron nitride (h-BN) monolayer}

As a starting point for further investigations of the elastic moduli of doped graphene, 
we present these quantities for pristine graphene and h-BN. 
The Young's, bulk, and shear moduli, and also Poisson's ratios for undoped systems
are gathered in Table~\ref{tab:tab1}.

All theoretical results obtained by us and others are very close. There is also a good 
agreement with experimental data for Young's and bulk moduli. 
Discrepancy between theory and experiment is for shear moduli, where all theoretical values
are larger. 

\subsection{Elastic properties of graphene monolayer functionalized 
with boron and nitrogen atoms}

\begin{figure*}
\includegraphics[width=0.95\textwidth]{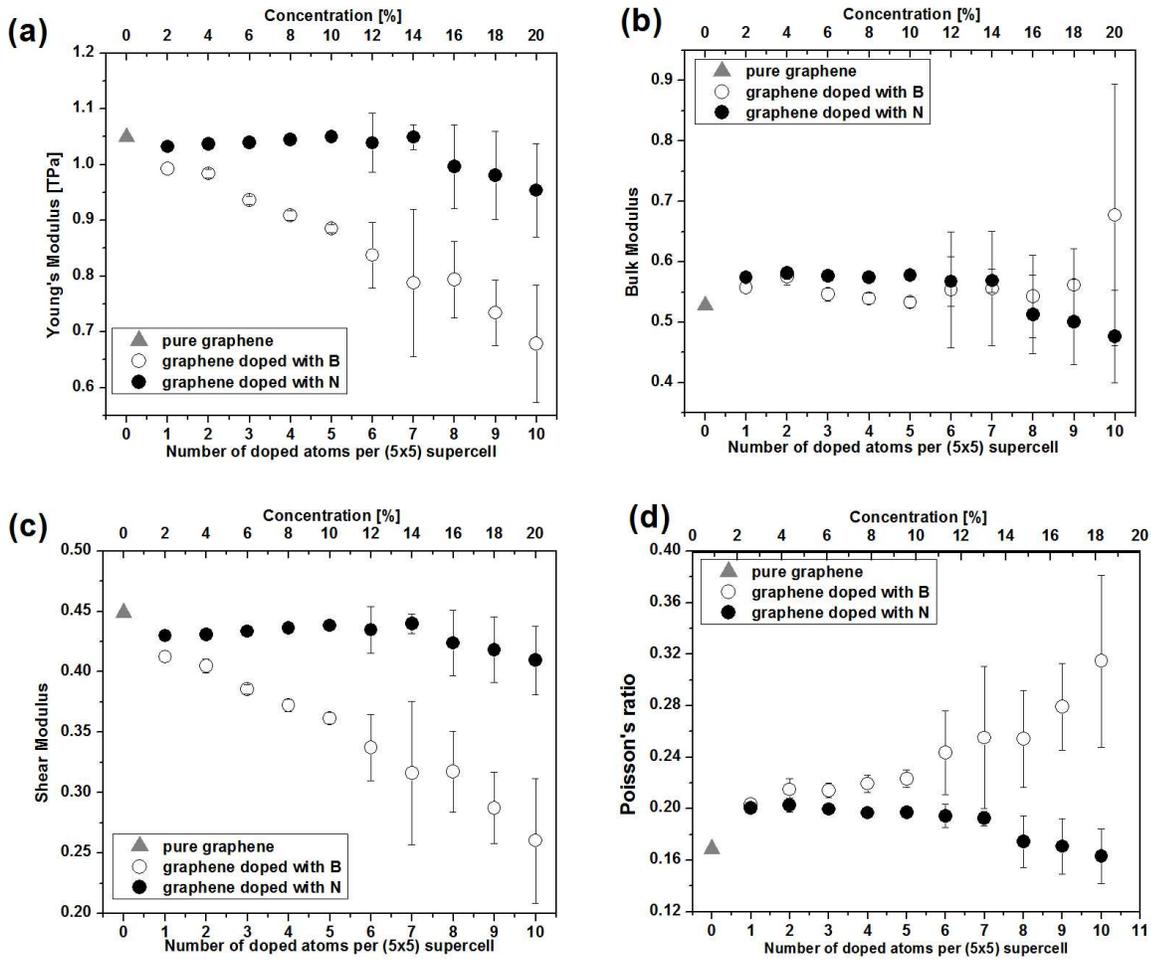}
\caption{\label{fig:elastyczne}
Young's (a), bulk (b) and shear (c) moduli and Poisson's ratio (d)
of graphene functionalized with B (empty circles) and N (filled circles)
as a function of the number of substituents per supercell and its percentage
concentration (top x-axis). Symbols mean the averaged values obtained from different
dopants arrangements in various supercells. 
The "error bars" are used to visualize the statistical spreads due to disorder.}
\end{figure*}

We start with the presentation of Young's, bulk and shear moduli and Poisson's ratio 
for N- and B-doped graphene, see Figure~\ref{fig:elastyczne}.

Firstly, Young's modulus decreases with the increasing number of dopants,
for both types of substituents, as shown in Figure~\ref{fig:elastyczne}(a).
However, the reduction of this quantity in comparison to  pure graphene is smaller 
for N-doped than for B-doped  graphene. At the same time, the analysis of 
the geometric structure shows that boron  induce more changes into the 
graphene layer than  nitrogen atoms.
For concentrations smaller than 14\%, one can see that Young's modulus 
of NG is almost unchanged in comparison to the undoped structure. 
Our observations are in good agreement with Mortazavi's works,
in which substituted nitrogen atoms maintain the stiffness
of the graphene layer up to the concentration of 6\% \cite{mortazaviN},
whereas 4\% of boron atoms reduce tensile strength by approximately 8\% 
in comparison to pure graphene\cite{mortazaviB}. 

Starting from the concentration of 12\%, the statistical spreads (depicted as error bars) 
for Young's modulus increase (Fig.~\ref{fig:elastyczne} (a)),
which means that the particular arrangement 
of dopants in graphene lattice starts to play an important role as far as stiffness 
of functionalized systems is concerned.  
To visualize this, see Figure~\ref{fig:drho} which displays four example 
structures of graphene doped with nitrogen and boron at the highest considered 
concentration of 20\%. For this concentration, the differences between Young's moduli 
of symmetrically nonequivalent configurations are equal to 0.231 TPa for NG 
and 0.315 TPa for BG, respectively. 
Figure~\ref{fig:drho} (a) and (c) refer to the case of the smallest Young's modulus, 
whereas (b) and (d) to the case of the highest value of this elastic constant 
observed in the investigated systems. 
NG depicted in Figure~\ref{fig:drho}(d) is characterized by  Young's modulus higher 
than observed for the pure graphene.
The stiffest structures of boron-doped graphene are those in which substituent atoms are 
aligned in the chains perpendicular to the direction of the applied tensile strain. 
In case of nitrogen-doped graphene, the structures characterized by the highest value of Young's 
modulus are those in which individual nitrogen atoms are surrounded by carbons, including 
the first and the second nearest neighbours. 
In our previous investigations of stability of boron and nitrogen doped graphene,\cite{arxiv}
we have shown that the configuration in which dopants are more clustered is energetically 
 less preferable than the configuration in which dopants are separated by carbon atoms.  
It was shown by Yuge\cite{yuge} that formation of the B-B and N-N bonds is disfavored
in comparison to forming of the B-C or N-C bonds. 
The experimental data\cite{zhao,wei} confirmed that N atoms are incorporated into graphene layer as pointlike dopants rather than clusters of dopant. Hanafusa et al.\cite{hanafusa} demonstrated that in B-doped graphite synthetised using diffusion method the substituent B atoms prefer not to have neighboring B atoms in the same hexagonal ring. Similar observation, but in B-doped graphene, was noted by Panchakarla et al.\cite{panchakarla2009}.

There are also  differences in the charge densities between particular arrangements 
of substituted atoms, clearly seen for higher concentrations.
For a better comparison, near all structures in Figure~\ref{fig:drho}, we have marked 
the participation of positive values (in \%) of the difference between the  valence pseudocharge 
density and a superposition of the atomic valence-pseudocharge densities at presented 
cross-sections of the substituted systems\footnote{The  cross-sections have been chosen in such 
a way that they are in planes parallel to the graphene lattice with the perpendicular coordinate 
chosen so that it is an average position of all distorted atoms in the supercell. }.
In the case of BG, the stiffer systems have higher concentration of 
positive charges in the cross-sections. 
Whereas in NG systems, Young's modulus is higher for  
these structures where higher electron density is observed.

\begin{figure}
\includegraphics[width=0.49\textwidth]{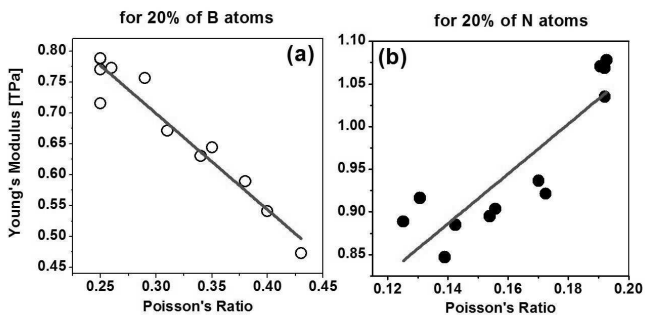}
\caption{\label{fig:ratio}
The relation  between Young's modulus and Poisson's ratio for all
the symmetrically nonequivalent configurations at 20\% concentration
of boron (a) and  nitrogen (b) atoms. 
Symbols mean values obtained from disordered samples.}
\end{figure}

Further, in Figure~\ref{fig:elastyczne}(b) and (c), 
we  present results for bulk and shear  moduli.
The magnitude of which can be easily calculated using Young's modulus and Poisson's ratio,
employing formulas ~\ref{ebulk} and ~\ref{eshear}, respectively.
For concentration of dopants up to 16\%, the bulk modulus does
not reflect a significant dependency on boron or nitrogen doping and its average value is
almost unchanged in comparison to the pure graphene. 
In contrast for higher concentrations, the average bulk modulus starts
to increase for BG and decrease for NG with the increasing number of substituents.
For the dopant concentration of 12\% and higher, 
the difference between symmetrically nonequivalent
configurations starts to be pronounced.

Shear modulus as a function of dopants, depicted in Fig.~\ref{fig:elastyczne}(c),
is very similar to the Young's  modulus (Figure~\ref{fig:elastyczne}(a)).
Shear deformations, which play an important role in wrinkling and rippling behavior of
graphene,\cite{katsnelson,liu2012} could be, like stiffness,  related to 
the differences in local charge densities of doped systems.
Particular arrangement of dopants in graphene layer, 
for concentration higher than 10\%, affects the charge carrier scattering and 
the electron mobility.
Shear modulus determines the resonance frequency of vibration modes 
involving torsional strains which, in contrast to the flexural strains, 
are not involved in thermoelastic loss.\cite{houston}
It is important to note, that the NG-based devices are advantageous over the BG-based devices 
if one is interested in the mechanical quality factors.

The most striking result, which distinguishes properties of the n- and p-type doped graphene,
shows up in Figure~\ref{fig:elastyczne}(d) which presents Poisson's ratio. 
This elastic property increases with the growing number of B atoms incorporated into 
the graphene layer and decreases with the increasing number of N atoms.
Graphene layers doped with 20\% of N atoms posses average Poisson's ratio almost the same 
as pure graphene (0.163). In contrast for 20\% of B atoms,  
this quantity is even higher than for h-BN (0.315). 
This ratio, which describes how easily the system is deformed in the direction perpendicular
to the applied load, allows to judge the difference between boron and nitrogen doped graphene
in this respect. 
For the same applied  stretching force,
the B-doped graphene will contract in transverse direction much more than N-doped graphene. 
For a better comparison of  the differences between both systems, we present plots of Young's 
modulus vs Poisson's ratio for the highest considered concentration 
in  Figure~\ref{fig:ratio}(a) and (b). 
Fitting a linear function to each case (using the least squares method), 
highlights two opposite tendencies:  the slope is negative for the boron-doped graphene, whereas for the nitrogen-doped samples, this slope  is positive.

\section{\label{sec:level4}Conclusions}

We present extensive and systematic studies of the elastic properties of B- and N-doped graphene.
We investigated structures with  2-20$\%$ of dopants in  various symmetrically nonequivalent configurations.
Our DFT  studies  include, essential for creation of graphene-based devices, comparison for two type of dopants
with respect to the concentration and particular arrangement of impurities.
The structure and related electronic properties of doped graphene have been investigated
on the basis of the simulated STM images and DRHO presenting samples morphology.
We demonstrated that the B-doped graphene exhibits much larger morphological changes compared
to the pure graphene than the N-doped samples. 

Our studies also provide valuable  predictions for Young's, shear and bulk moduli and Poisson's ratio
 of doped graphene, and evidence clear chemical trends, 
 shedding light on physical mechanisms governing these effects. 
From the analysis of disorder and aggregation,
we observe a general rule - aligning B atoms in the chains perpendicular to 
the applied tensile strain or separating N-dopants among themselves by two shells of C atoms, 
builds samples with the highest Young's and shear moduli.
We claim that some way of incorporating N atoms in the graphene layer could  
even strengthen the doped system in comparison to the pure graphene. Noticeable, all B-type 
doping reduces the stiffness.  
Bulk modulus and the Poisson's ratio for the concentrations 
greater than 16\% and 12\%, respectively, exhibit opposite trends:
the B atoms improve the elastic constants while the N atoms worsen them. 
In the case of N-doped graphene, all of the considered elastic properties remain
almost unchanged for concentrations smaller than 14\%. 

We have found that the elastic properties are strongly related to the electronic properties 
of doped systems, especially for heavily doped samples. Particular arrangement of dopants 
in graphene layer and the related local charge-densities play important 
role in mechanical quality factors in BG and NG. 
Each type of doping introduces different changes with respect to all considered properties, 
and these differences cannot be neglected. 

The conclusions drawn in this work are important for the design of flexible electronics 
and nanocomposites or other applications in fields such as fuel cells  or hydrogen storage,
employing graphene layers. 

\section{\label{sec:level5}Acknowledgments}
The authors acknowledge financial support of the SiCMAT Project financed under the European Founds 
for Regional Development (Contract No. UDA-POIG.01.03.01-14-155/09). 
We thank also PL-Grid Infrastructure and Interdisciplinary Centre for Mathematical and 
Computational Modeling of University of Warsaw (Grant No. G47-5 and G47-16) for providing 
computer facilities.

\providecommand*\mcitethebibliography{\thebibliography}
\csname @ifundefined\endcsname{endmcitethebibliography}
  {\let\endmcitethebibliography\endthebibliography}{}

\end{document}